\documentclass[twocolumn]{elsart3p}

\usepackage{epsfig}

 \usepackage{lineno}

\begin{document}

\begin{frontmatter}

\title{Detailed investigation on the possibility of using EJ-299-33A plastic scintillator for fast neutron spectroscopy in large scale experiments}

\author[lable1]{Pratap Roy\corauthref{cor}},
\corauth[cor]{Corresponding author.}
\ead{pratap\_presi@yahoo.co.in}
\author[lable1,lable2]{K. Banerjee},
\author[lable1]{A. K. Saha},
\author[lable1,lable2]{C. Bhattacharya},
\author[lable1]{J. K. Meena},
\author[lable1]{P. Bhaskar},
\author[lable1,lable2]{S. Mukhopadhyay},
\author[lable1]{S. Bhattacharya}.

\address[lable1]{Variable Energy Cyclotron Centre, 1/AF-Bidhannagar, Kolkata-700064, India.}
\address[lable2]{Homi Bhabha National Institute(HBNI), Training School Complex, Anushaktinagar, Mumbai~-~400094, India.}

\begin{abstract}
Detailed characterization of the newly available plastic scintillator (EJ-299-33A) having the pulse shape discrimination (PSD) property has been carried out in case of a large-sized detector (5 in.$\times$5 in.). The pulse height response of the scintillator for nearly mono-energetic neutrons has been reported in case of neutron energies E$_n$ =3, 6 and 9 MeV. Important properties (figure-of-merit (FOM), time resolution, detection efficiency) of the detector has been compared with a commonly used liquid organic scintillator based detector of the same size coupled to the same PMT for uniformity in comparison. The results show that the plastic scintillator detector has about 12$\%$ better time resolution. However, the FOM and detection efficiency were found to be lower than that of the liquid scintillator detector by 40~-~50$\%$ and $\sim $25$\%$, respectively. The possibility of using the new plastic scintillator in large-scale nuclear physics experiments has been pointed out.

\end{abstract}

\end{frontmatter}
\section{Introduction}

A number of large-scale neutron detector arrays for fast neutron spectroscopy have been developed in recent times at various parts of the world~\cite{Trino,NAND,Rout,LANSCE,MONA}. A 50-element liquid scintillator based neutron detector array is also being constructed at VECC. Additionally, a part of the MONSTER array~\cite{MONSTER} for FAIR~\cite{FAIR} is being developed at VECC at the same time. Several exciting physics issues like $\beta $-delayed neutron emission, nuclear fission dynamics, production of super-heavy elements (SHE) can be addressed using these modular neutron detector arrays. Neutron time of flight (n$\_$TOF) arrays would also be essential to study the structure of the neutron-rich nuclei to be produced in upcoming big radioactive ion-beam (RIB) facilities like FRIB at MSU, USA, RIBF at RIKEN, Japan and FAIR at Germany.  In most of these arrays organic liquid scintillators (LS) have been the preferred choice as the detection medium because of the high detection efficiency, fast timing characteristic and most importantly due to their excellent ability to discriminate between the neutron and $\gamma$-ray events. However, many of the liquid scintillators suffer from the problems like toxicity, low flash point, and chemical hazards. These problems can raise severe safety concerns particularly when a considerable number of detectors are involved. Another problem with large liquid scintillator cells often encountered by the regular users is the problem of leakage of the liquid through the micro-leaks present around the joints. The leakage can result in the formation of undesired bubbles inside the detector cells which may degrade the quality of the pulses. All the problems associated with liquid scintillators can be avoided by the use of plastic scintillators (PS) which can also serve as a useful detection medium for the neutron. However, unlike liquid scintillators plastic scintillators, until recently, lacked the pulse shape discrimination (PSD) property which is needed to discriminate between the neutron and $\gamma$-ray events. The long-lasting desire to have plastic scintillators with good PSD properties may seem to be achieved with the commercial release of a new type of plastic scintillator by Eljen Technology~\cite{Eljen} named EJ-299-33, which enables the separation of $\gamma$ and fast neutron signals on the basis of their timing characteristics using conventional PSD electronics techniques. The possibility of synthesizing a plastic scintillator with efficient neutron-$\gamma $ discrimination ability using a combination of the polyvinyltoluene (PVT) matrix loaded with traditional fluorescent compounds was revealed by Natalia {\it et al.}~\cite{Natalia}. The first demonstration on the PSD capabilities of the new plastic scintillator was presented by Pozzi {\it et al.}~\cite{Pozzi} using a 5.08-cm diameter by 5.08-cm thick detector. In another recent article by Cester {\it et al.}~\cite{Cester} the characteristics of a cylindrical 5 cm$\times $ 5 cm EJ-299-33 detector has been reported, and a comparison has been made with other conventional scintillators ({\it e.g.}, EJ~301 and EJ~309). The radioluminescent characteristics of a 5 cm$\times $ 5 cm EJ-299-33 plastic scintillator have also been reported by Nyibule {\it et al.}~\cite{Nyibule}. All the above reports showed promising results; however, they were limited to rather limited sized detectors. For the use in large-scale nuclear physics experiments, one would like to have large sized detectors mainly to increase the detection efficiency. An improved version (EJ-299-33A) of the original PSD plastic scintillator has been introduced recently that embodies a significant color improvement resulting in improved transparency, which is particularly crucial for fabricating larger scintillators. This material could serve as a potential replacement of the conventional liquid scintillators in large-scale neutron detector arrays provided its performance is comparable to that of LS detectors. In this scenario, it is tempting to measure important characteristics of the new plastic scintillator and compare it with the traditional liquid scintillator-based detector. \\
In this paper, we report the (i) PSD property, (ii) pulse height response for nearly mono-energetic neutrons, (iii) timing characteristics, and (iv) energy-dependent neutron detection efficiency in case of a 5 in.$\times$5 in. EJ-299-33A plastic scintillator detector. The same quantities have also been measured in case of a similar liquid scintillator (BC501A) based detector in the same experimental conditions for making the most meaningful comparison.    

\section{Experimental details}
The EJ-299-33A plastic scintillator detector manufactured by SCIONIX, Holland consists of a 5 in. (dia.)$\times$5 in. (length) cylindrical detector cell coupled to a 5 in. R4144 HAMAMATSU photomultiplier tube (PMT). A similar detector based on one of the most commonly used liquid scintillator, BC501A has been fabricated at VECC and coupled to the same photomultiplier tube. The pulse height (PH) and the pulse shape discrimination property of the neutron detectors have been measured using a single width dual channel NIM based integrated electronics module~\cite{Venkat}. The dynode and anode outputs from PMT are fed into the integrated electronics module which consists of a pre-amplifier, shaper amplifier, constant fraction discriminator (CFD), PSD and time to amplitude converter (TAC) circuits built in it. This module adopts zero cross-over (ZCO) technique for neutron-$\gamma $ pulse shape discrimination. In the ZCO method, the anode signal from the fast PMT is equally divided through the resistive signal splitter and coupled to (i) CFD section and (ii) PSD section. The signal applied to zero cross-over amplifier section is suitably differentiated and integrated ($\sim$300 ns) to realize a semi-Gaussian bipolar signal. Thus generated bipolar signals have different zero cross-over time (ZCT), which is utilized for recording n-$\gamma $ separation. The energy (pulse height) was measured by connecting the last dynode output of the PMT to a charge sensitive pre-amplifier and a shaping amplifier with shaping time of $\sim $1 $\mu $s.\\  
\begin{figure}
\begin{center}
\includegraphics[scale=0.5]{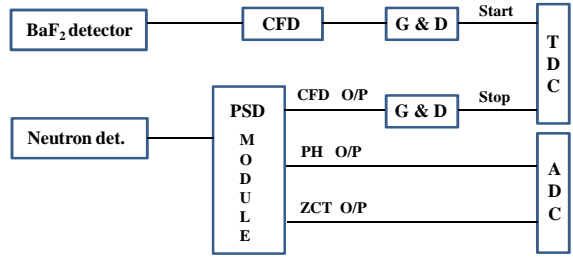}
\caption{\label{energy} Block diagram of the experimental setup.}
\end{center}
\end{figure}
The time resolution of the scintillator detectors was measured by measuring the time distribution of the coincidence $\gamma $-rays emitted from a $^{60}$Co source with reference to a fast BaF$_2$ (Dimension: tapered, front dia. 2.5 cm, back dia. 2 cm, length $\sim $2 cm, time resolution: $\sim $310 ps) detector. The $^{60}$Co source was placed in between the BaF$_2$ and the scintillator detector where the separation between them was about $\sim $30 cm. The time resolution of the BaF$_2$ detector was measured using two identical detectors and was corrected from the measured overall time resolution as per the following relation
\begin{equation}
(FWHM_{tot})^2=(FWHM_{det1})^2+(FWHM_{det2})^2
\end{equation}
For the PH response measurements a time of flight (TOF) setup was established where the fast BaF$_2$ detector was used as the reference to generate the START trigger for the TOF measurement. The response functions of nearly mono-energetic neutrons  below 10 MeV have been extracted from a corresponding neutron energy spectrum of $^{252}$Cf neutron source measured through the TOF technique. The $^{252}$Cf source ($\sim $35 $\mu $Ci) was placed just in front of the BaF$_2$ detector (START detector), and the neutron detector (STOP detector) was placed at a distance of 150 cm from the source. A block diagram of the experimental setup has been shown in Fig. 1. 
\begin{figure}
\begin{center}
\includegraphics[scale=0.5]{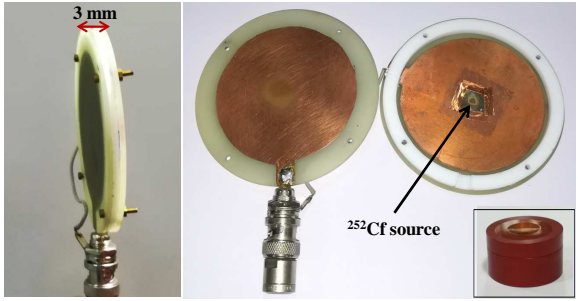}
\caption{\label{energy} Photograph of the fission chamber. The $^{252}$Cf source (small circular spot in the picture) is mounted on one of the electrodes. The original metallic holder of the $^{252}$Cf source is shown in the inset.}
\end{center}
\end{figure}
The prompt $\gamma $-rays emitted from the $^{252}$Cf source have been detected by the BaF$_2$ detector and the fast (sub-nanosecond) component of the detected signals have been used to generate the time reference for the TOF measurement. The TOF spectra of neutrons have been generated from the measured time difference between the BaF$_2$ signal and the neutron detector signal. While the neutron energies were measured from the time-of-flight, the n-$\gamma $ separation was achieved by both TOF and PSD measurements. A two-dimensional correlation plot between the measured ZCT and TOF has been generated to discriminate between the neutron and $\gamma $-ray events.  The mono-energetic neutron response has been extracted from the measured continuous energy neutron spectrum by putting appropriate TOF-gate in the ZCT vs. TOF spectrum. \\
\begin{figure}
\begin{center}
\vspace{0.3cm}
\includegraphics[height= 5.5cm, width=8.0 cm]{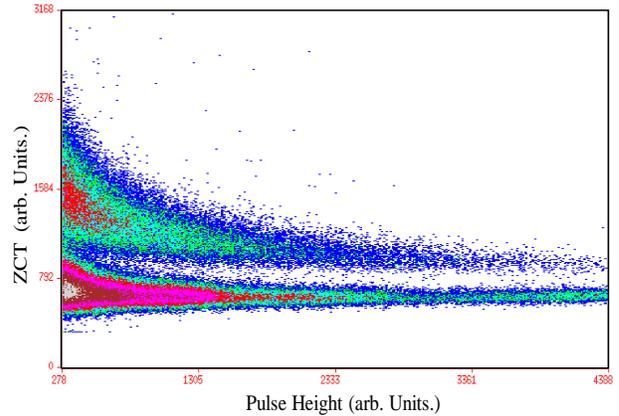}
\caption{\label{time} Typical two-dimensional ZCT vs. PH plot at a threshold of 300 keVee for the EJ-299-33A detector obtained using Am-Be neutron source.}
\end{center}
\end{figure}

For the efficiency measurement the BaF$_2$ detector in Fig.~1 was replaced by a small fission detector~\cite{Vishal} which detects the fission fragments emitted by $^{252}$Cf. The fission chamber (FC) consists of two parallel copper coated G10 plates (circular, dia. $\approx $8 cm) separated by a 3~mm thick Teflon ring (Fig.~2). The two plates are connected to a BNC connector to apply high voltage as well as taking the detector signal out. The main advantage of this detector is that it uses air as the detection medium and need not be operated in vacuum. The $^{252}$Cf source was mounted within the detector on one of the electrodes. This detector was operated in air with bias $\sim$600 Volt applied between the two electrodes. As earlier the neutron energy was measured using the TOF technique, where the start signal was taken from the fission trigger detector, and the stop signal was taken from the neutron detector. The neutron detector to fission detector (source) distance was kept at 2.0 m during the measurement. The TDC calibration has been checked several times during the experiment using a precision time calibrator~\cite{Ortec}. The prompt $\gamma $-peak in the TOF spectrum has been used as the reference time for neutron time of flight analysis. Data from different detectors were recorded using a VME based data acquisition system~\cite{LAMPS} on event-by-event basis.
\begin{figure}
\begin{center}
\includegraphics[scale=0.75]{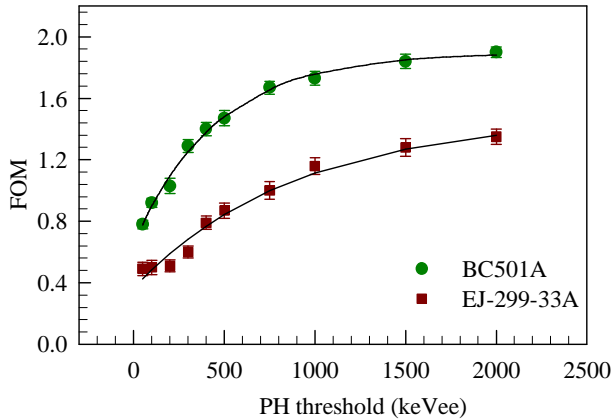}
\caption{\label{setup} Figure of merit for different pulse height threshold. Continuous lines are average fits.}
\end{center}
\end{figure}

\section{Results and discussions}
\subsection{Pulse shape discrimination}     
The energy calibration of the neutron detectors has been carried out from the measured Compton distribution of the known $\gamma $-ray sources ($^{137}$Cs, $^{22}$Na and $^{241}$Am-$^9$Be). In determining the position of the Compton edge, the prescription of Ref.~\cite{Dietze} was followed. The energy calibration was found to be highly linear in the measured energy range (up to 4.4 MeV). The pulse shape discrimination property was investigated using a $^{241}$Am-$^9$Be neutron source. Fig.~3 shows a typical ZCT vs. PH two-dimensional plot at a pulse height threshold of 300 keVee. In order to characterize the n-$\gamma $ discrimination ability, the figure of merit (FOM) was defined in a conventional manner,
\begin{equation}
FOM~=~\frac{\Delta }{\delta_n + \delta_g }
\end{equation}

where $\Delta $ is the separation between the centroids of the neutron and $\gamma $ peaks, and $\delta_n$ and $\delta_g$ are the full-width at half-maximum (FWHM) of the neutron and $\gamma$ peaks, respectively. The CFD walk parameter was adjusted to obtain the optimum value of the FOM. Variation of the FOM with the increase in the PH threshold is shown in Fig.~4 for both the PS and LS detectors. It can be seen that, although reasonable separation between the neutron and $\gamma $-rays is obtained (Fig.~3) in case of the new PS detector, the figure of merit is consistently lower (Fig.~4) than that of the similar LS detector. For example, the FOM of the PS detector is about $\sim $46$\%$ and $\sim $67$\%$ to that of the LS detector at a PH threshold of 300 and 1000 keVee, respectively. The FOM of the measured EJ-299-33A detector is comparable to that of the reported values in case of a smaller sized EJ-299-33 detector~\cite{Cester}. This can be considered as a reasonable improvement on the PSD property of the new plastic scintillator as the n-$\gamma $ discrimination property is expected to degrade with the increase in detector dimension~\cite{kaushik}.  
\begin{figure}
\begin{center}
\vspace{0.3cm}
\includegraphics[scale=0.75]{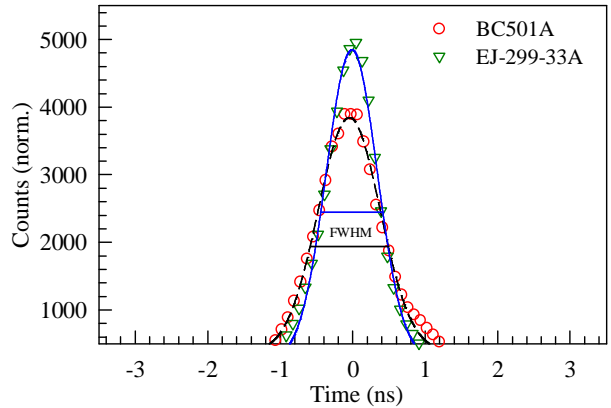}
\caption{\label{time} Time distribution of the coincidence $\gamma$-rays between the reference BaF$_2$ detector and BC501A (open circles), EJ-299-33A (open triangles) detectors. The lines represent the Gaussian fits.}
\end{center}
\end{figure}

\subsection{Time resolution} 
The timing characteristics are particularly important when the energy is measured through the time of flight (TOF) technique. The measured time distributions of the coincidence $\gamma $-rays are shown in Fig.~5. The time resolution of the scintillator detectors were determined using Eq.~(1) and found to be 960~$\pm$~40 ps and 1100~$\pm$~50 ps for the EJ-299-33A and BC501A detectors, respectively. It is found that the PS detector is about 12$\%$ faster in time resolution compared to LS at a pulse height threshold of 100 keVee making it a very good choice for fast timing applications. 
\begin{figure}
\begin{center}
\vspace{0.3cm}
\includegraphics[height= 6.0cm, width=8.0 cm]{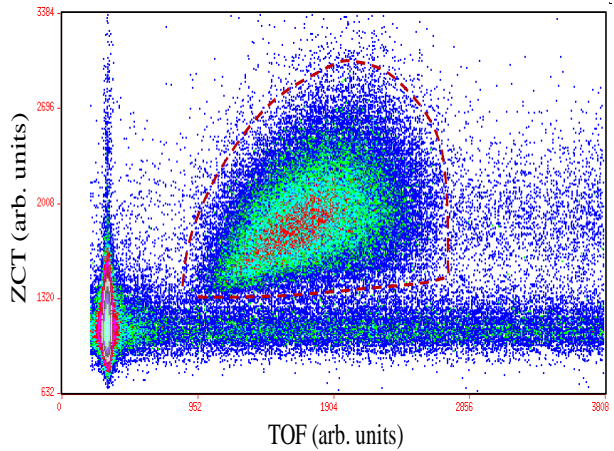}
\caption{\label{time} A typical two-dimensional ZCT vs. TOF plot in case of the EJ-299-33A detector showing the complete separation between the $\gamma $-ray and neutron events (shown within the dashed (red) contour).}
\end{center}
\end{figure}
\begin{table}

\caption[]{\label{tab:nucl} Average pulse-height corresponding to different neutron energies.}
\vskip 0.5cm
\begin{center}		
\begin{tabular}{|c|c|c|}
\hline
{\bf Scintillator}     & {\bf Mean neutron}  & {\bf Average pulse}  \\
 &     {\bf energy (MeVee)}  &          {\bf height $<H>$(MeVee)}       \\
 \hline
 \hline
 &     3~$\pm$~0.1   &    0.32  \\
\hline
{ EJ-299-33A} &     6~$\pm$~0.3    &  0.69 	  \\
\hline
   &  9~$\pm$~0.7    &  0.99    	\\
\hline
\hline
&     3~$\pm$~0.1   &    0.38  \\
\hline
{ BC~501A} &     6~$\pm$~0.3    &  0.82 	  \\
\hline
   &  9~$\pm$~0.7    &  1.17    	\\
\hline
\end{tabular}
\end{center}		
\end{table}
\begin{figure}
\begin{center}
\vspace{0.3cm}
\includegraphics[scale=0.73]{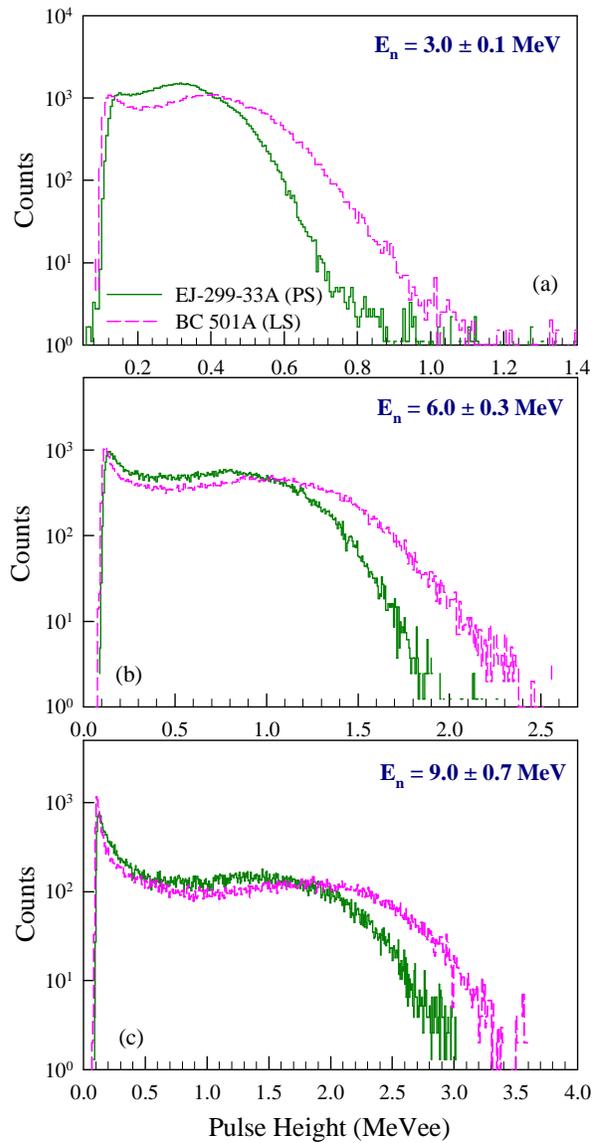}
\caption{\label{time} Pulse-height spectra for neutrons with energies 3, 6 and 9 MeV measured using a $^{252}$Cf source. The counts have been normalized to make the area under the curves same}.
\end{center}
\end{figure}
\subsection{Pulse height response}
The pulse height response of the EJ-299-33A detector was measured for the first time for three mean neutron energies E$_n$~=~3 ($\pm $0.1), 6 ($\pm $0.3), and 9 ($\pm $0.7) MeV. The neutron energies were selected by putting suitable time-of-flight gates in the two-dimensional ZCT vs. TOF plot which is shown in Fig.~6. It can be seen from Fig.~6 that neutron and $\gamma$-ray events are nicely discriminated, and there is hardly any overlap between the two. The pulse height (H) distribution of the neutron events (shown by the red-dashed contour in Fig.~6) was extracted within a given TOF window. The pulse-height response for three neutron energies is displayed in Fig.~7. The response of the BC501A detector has also been shown in the same figure for comparison. The average pulse-heights ($<H>$) corresponding to different neutron energies are given in Table~1. The average pulse heights are calculated from the measured PH distributions using the standard relation,
\begin{figure}
\begin{center}
\vspace{0.3cm}
\includegraphics[scale=0.75]{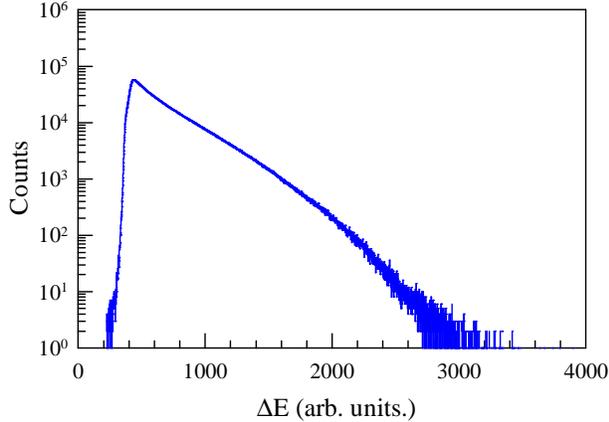}
\caption{\label{time} A typical fission fragment energy loss spectrum measured through the fission chamber. }
\end{center}
\end{figure}
\begin{equation}
<H>~ = \frac{\int{H~N(H)~dH}}{\int{N(H)~dH}}
\end{equation}
where N(H) is the number of counts for a given pulse height H. It can be seen from Fig.~7 and Table~1 that the average pulse-height of the EJ-299-33A detector is consistently lower than the corresponding liquid scintillator detector. It establishes the fact that the EJ-299-33A plastic scintillator is characterized by a lower light output at given energy than the liquid scintillator. The difference in the average pulse-height is found to be around ~18$\%$ at all the measured energies.

\subsection{Energy-dependent efficiency}
The $^{252}$Cf neutron energy spectra were generated from the measured TOF spectra which were obtained by taking the projection of the two-dimensional ZCO vs. TOF scatter plot onto the TOF axis. The detection efficiency is defined by the number of detected neutrons divided by the number of neutrons incident on the detector as a function of neutron energy. The energy distribution of the incident neutrons were determined by expected energy distribution for $^{252}$Cf given by~\cite{Madland},
\begin{equation}
N(E)= \frac{2\sqrt{E}exp(-E/T)}{\sqrt{\pi }(T)^{3/2}}
\end{equation}
The total number of incident neutrons was determined from the total number of fission events determined by taking the area of the fission fragment energy loss ($\Delta E$) spectra (Fig.~8) measured through the small fission chamber. It may be noted here that the time resolution of fission chamber ($\sim $3 ns) is not the excellent one, and it is generally not recommended for precise TOF measurements. However, in the present case, it was used to keep the measurement simple without losing the essential information. The uncertainty ($\Delta $E) in neutron energy (E) was estimated using the relation,
\begin{equation}
\frac{\Delta E}{E}= \sqrt{({\frac{2\Delta \tau }{t}})^{2}+{(\frac{2\Delta L}{L}})^{2}}
\end{equation}
\begin{figure}
\begin{center}
\vspace{0.3cm}
\includegraphics[scale=0.5]{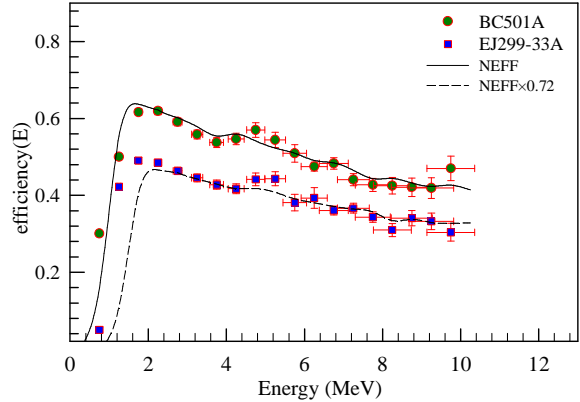}
\caption{\label{time} The measured efficiency at a PH threshold of 100 keVee. Lines are the NEFF predictions. }
\end{center}
\end{figure}
where $\Delta \tau $ is the time resolution, $t$ is neutron flight time, $L$ is neutron mean flight path, $\Delta L$ is the flight path spread due to the detector size. In the present case the maximum uncertainty in neutron energy (corresponding to the highest E) was found to be around $\sim $15$\%$. 
The measured efficiencies at a pulse-height threshold of 100 keVee have been shown in Fig.~9. The uncertainty in the energy measurement determined through the Eq.~(4) has been displayed by error bars in Fig.~9. The measured efficiency for the EJ-299-33A PS detector was found to be about $\sim $25$\%$ lower (at 2 MeV) than the BC501A LS detector. It may be pointed out here that the plastic scintillator has higher density (1.08 gm/cm$^3$) compared to the liquid scintillator (0.875 gm/cm$^3$). The higher density would result in enhanced detection efficiency. However, the plastic scintillator has lower H:C ratio and characterized by lower light output than the liquid scintillator. Both these effects will reduce the detection efficiency. The experimentally measured efficiencies were also compared with the Monte-Carlo based simulation carried out using the NEFF~\cite{Dietze2} code. It can be seen from Fig.~9 that the measured efficiency for the LS detector is in good agreement with the NEFF prediction. One can see a small bump in the efficiency around 4.5 MeV. This may be because of the opening up of $^{12}C(n,n')$$^{12}C^*$ reaction channel after 4.4 MeV (1$^{st}$ excited state of $^{12}$C). In the NEFF calculation for the PS detector proper density and H:C ratio has been incorporated, however, the light output was used as same as that of the LS. Finally, the predicted efficiency was scaled by a reduction factor to match the measured data. It was found that a scaling factor of 0.72 reproduces the measured data quite reasonably. Just from the PH measurement (Sec~3.3) one would expect a reduction in the efficiency of the PS detector by 18~-~20$\%$. It would be interesting to carry out detailed simulation of the efficiency and PH response for the PS detector using appropriate energy-dependent light output functions and other factors like density, H:C ratio.     

\section{Summary and conclusion}

Pulse shape discrimination, time resolution and efficiency of an EJ-299-33A plastic scintillator based detector of size 5 in.$\times $ 5 in. have been measured exploring the possibility of its use in large-scale nuclear physics experiments. All properties of the plastic detector have been compared with a similar liquid scintillator (BC501A) detector with the same PMT (HAMAMATSU R4144). The new plastic detector has lower figure-of-merit (about 46$\%$ of the LS at 300 keVee) for n-$\gamma $ separation particularly for low pulse-height events. However, as shown in the present work, reasonable separation between the neutron and $\gamma $-ray events can be achieved by combining techniques like PSD and TOF. Compared to the LS, the plastic scintillator is characterized by a lower light output, which also reduces the detection efficiency by $\sim $25$\%$ (at E$_n$~= 2 MeV). The timing characteristic of the plastic scintillator was found to be slightly better ($\sim $12$\%$ faster) than the corresponding liquid scintillator. The overall characteristics of the newly available plastic scintillator certainly make it suitable for fast neutron measurements. However, still, there are scopes for further improvements which may make this material comparable to the liquid scintillators. In fact, the very recent introduction of the third generation of the PSD plastic scintillator (EJ-276) which claims to have even better PSD capability~\cite{Zaitseva}; could make the new material an excellent alternative to the commonly used liquid scintillators in nuclear physics experiments. It will be interesting to extend the measurements carried out in the present work to the latest PSD plastic scintillator with large dimensions. 

\section{Acknowledgment}
The authors would like to thank Dr. B. K. Nayak of NPD, BARC for providing essential inputs for the efficiency measurement using the small fission chamber.



\begin{thebibliography}{99}

\bibitem{Trino} T. Martinez {\it et al.}, Nuclear Data Sheets {\bf 120} (2014) 78.
\bibitem{NAND} NAND, $http://www.iuac.res.in/research/np/nand/nand\_main.html$.
\bibitem{Rout} P. C. Rout {\it et al.}, Proceedings of the DAE Symp. on Nucl. Phys. {\bf 62} (2017) 1028.
\bibitem{LANSCE} B. A. Perdue {\it et al.}, IEEE Transactions on Nucl. Sci. {\bf 60}, (2013) 879.
\bibitem{MONA} MONA, $http://mona.wabash.edu/html/default/home.html$.
\bibitem{MONSTER} MONSTER TDR, $https://fair-center.eu/for-users/experiments/nustar/documents/technical-design-reports.html$.
\bibitem{FAIR} FAIR, $http://www.fair-center.eu$.
\bibitem{Eljen} Eljen Technology, ($http://www.eljentechnology.com$).
\bibitem{Natalia} Natalia Zaitseva {\it et al.}, Nucl. Instr. and Meth. {\bf A 668} (2012) 88.
\bibitem{Pozzi} S.A.Pozzi, M.M.Bourne, S.D.Clarke, Nucl. Instr. and Meth. {\bf A 723} (2013) 19.
\bibitem{Cester} D. Cester {\it et al.}, Nucl. Instr. and Meth. {\bf A 735} (2014) 202.
\bibitem{Nyibule} S. Nyibule {\it et al.}, Nucl. Instr. and Meth. {\bf A 728} (2013) 36.
\bibitem{Venkat} S. Venkataramanan {\it et al.},  Nucl. Instr. and Meth. {\bf A 596} (2008) 248.
\bibitem{Vishal} V. V. Desai {\it et al.}, Phys. Rev. {\bf C 92}, 014609 (2015).
\bibitem{Ortec} Time Calibrator $http://www.ortec-online.com/products/electronics/time-to-amplitude-converters-tac/462$.
\bibitem{LAMPS} LAMPS, $http://www.tifr.res.in/~pell/lamps.html$.
\bibitem {Dietze} G. Dietze and H. Klein, Nucl. Inst. Meth. {\bf 193}, (1982) 549.
\bibitem{kaushik} K. Banerjee {\it et al.}, Nucl. Instr. and Meth. {\bf A 608} (2009) 440.
\bibitem{Madland} D.G. Madland and J.R.Nix, Nucl. Sci. and Eng. {\bf 81}, 213 (1982).
\bibitem{Dietze2} G. Dietze, H. Klein, PTB-ND-22 Report, 1982.
\bibitem{Zaitseva} N.P. Zaitseva {\it et al.}, Nucl. Instr. and Meth. {\bf A 889} (2018) 97.


\end{thebibliography}
\end{document}